\documentclass[aps,prd,preprint,amsmath,citeautoscript,longbibliography,nofootinbib]{revtex4-1}
\usepackage[utf8]{inputenc}
\usepackage{bm}
\usepackage{amsmath,mathrsfs,amsfonts}
\usepackage{graphicx,afterpage}
\usepackage{subcaption}
\usepackage{amsmath,latexsym}
\usepackage{color,soul}
\usepackage{float}
\usepackage[section]{placeins}
\usepackage{silence}
\usepackage[hidelinks]{hyperref}
\begin{document}
\title{Consistency Conditions and the Derivation of Harmonic Structure in Einstein-Maxwell-dilaton Theory}
\author{Bardia H. Fahim}
\email{bardia.fahim@usask.ca}
\affiliation{Department of Physics and Engineering Physics, University of Saskatchewan, Saskatoon SK S7N 5E2, Canada}
\affiliation{School of Trades \& Technology,
Red Deer Polytechnic, Red Deer AB T4R 0V5, Canada}
\author{Joshua G. Fenwick}
\email{joshua.fenwick@usask.ca}
\affiliation{Department of Physics and Engineering Physics, University of Saskatchewan, Saskatoon SK S7N 5E2, Canada}
\date{\today}
\begin{abstract}

Many exact solutions of Einstein-Maxwell and Einstein-Maxwell-dilaton theory share a common structural pattern in which the metric functions are built from harmonic functions on a specified spatial base, often taken to be flat. We investigate this pattern by considering the Einstein-Maxwell-dilaton theory in arbitrary dimensions, where the dilaton field is non-trivially coupled to the Maxwell field and to a Liouville-type potential proportional to the cosmological parameter. Without imposing either the base geometry or the harmonic form of the metric function in advance, we show that for the generic branch, the field equations force the dilaton couplings to be equal, restrict the spatial base geometry to be Ricci flat, and make the metric function harmonic on this base. The resulting spacetime can then be written in terms of a conformal potential. We also consider a purely spatial branch, which instead leads to a distinct constraint on the coupling constants. These results provide a unified field-equation derivation of the harmonic behavior and conformal structure that appear in several classes of solutions, including multi-center geometries, cosmological solutions, and dynamical black holes. 
    
\end{abstract}

\maketitle
%\tableofcontents

%%%%%%%%%%%%%%%%%%%%%%%%%%%%%%%%%%%%%%%%%%   SECTION 1   %%%%%%%%%%%%%%%%%%%%%%%%%%%%%%%%%%%%%%%%%%%%%%%%%%%%%%%%
\newpage
\section{Introduction}
\label{sec:intro}

The construction of exact solutions to Einstein gravity and its extensions remains one of the central directions in gravitational physics. Such solutions provide valuable insight into the nonlinear structure of the field equations and play an essential role in understanding black holes, cosmological evolution, and higher dimensional theories motivated by string theory and dimensional compactification. In particular, Einstein-Maxwell-dilaton (EMD) theories have attracted considerable attention since the dilaton field appears naturally in low-energy string theory and provides a natural framework for studying charged gravitational systems with scalar field effects. EMD theory has been studied in a wide range of contexts, including Kaluza-Klein compactification \cite{gibbons1986black,gibbons1988black}, charged dilatonic solutions \cite{goldstein2010holography, poletti1995charged}, string-inspired black holes and branes \cite{horowitz1991black}, dynamical and radiating black holes \cite{aniceto2016radiating, hirschmann2018black}, and cosmic censorship \cite{yu2018cosmic}. 

There is a recurring structural pattern in many well-known Einstein-Maxwell and Einstein-Maxwell-dilaton solutions, in which the metric functions are governed by harmonic functions on the spatial base. This structure appears in different physical settings, including the Majumdar-Papapetrou and Hartle-Hawking multi-black-hole solutions \cite{majumdar1947class, papapetrou1945static, hartle1972solutions}, cosmological Einstein-Maxwell backgrounds \cite{kastor1993cosmological}, their dilatonic extensions \cite{shiraishi1993multicentered}, and EMD solutions with nontrivial spatial geometries \cite{ghezelbash2015cosmological,fahim2024exact}. In many of these cases, the harmonic function also controls a conformal scaling of the spatial metric. The repeated appearance of this pattern suggests that it is not merely a convenient ansatz, but may reflect a deeper consequence of the coupled gravitational and matter field equations.   

Motivated by this observation, in the present work we investigate this recurring pattern within the framework of EMD theory in arbitrary dimensions, where the dilaton is coupled to the Maxwell field and governed by a Liouville-type potential. Within the class of ansatzes considered here, we leave the spatial base geometry arbitrary and analyze the cases of mixed time-space dependence and purely spatial dependence of the metric function separately. We show that the field equations impose strong restrictions on the geometry of the resulting solutions. In the generic mixed branch, consistency forces the two dilaton couplings to coincide, restricts the spatial base to be Ricci flat, and requires the metric function to be harmonic on the base geometry. As a result, the solution can be written in terms of a single conformal potential. We also study the purely spatial branch, where the absence of mixed time-space dependence leads to a different restriction on the coupling constants. Within this general framework, several of these well-known solutions are recovered as particular limits, including the multi-center Einstein-Maxwell geometries and their EMD extensions. The result therefore provides a common field equation origin for the harmonic behaviour and conformal structure appearing across different Einstein-Maxwell and EMD theories.

The organization of this paper is as follows. In section \ref{sec2}, we introduce the EMD theory in $(N+1)$ dimensions, the corresponding field equations, and the ansatzes used throughout the paper. In section \ref{sec3}, we analyze the mixed time-space branch and derive the main consistency conditions leading to the conformal potential form of the solution. We also discuss the Einstein-Maxwell limit and the vanishing cosmological constant limit of the EMD theory. In section \ref{sec4} we study the purely spatial branch and the corresponding restrictions on the coupling constants and metric function. In section \ref{sec5} we relate the resulting framework to known Einstein-Maxwell and Einstein-Maxwell-dilaton solutions, including multi black holes, cosmological examples, and geometries with nontrivial spatial bases. We conclude the paper in section \ref{sec6} with a summary of the main results and comments on possible future directions.

\section{Einstein-Maxwell-dilaton theory in $(N+1)$ dimensions } \label{sec2}

We consider an $(N+1)$-dimensional Einstein-Maxwell-dilaton theory in which the dilaton field $\phi$ is non-minimally coupled to the Maxwell field and is governed by a Liouville-type potential. Throughout this paper, we use geometric units in which $c=1$ and $16\pi G=1$. The action for this theory is given by \cite{charmousis2009einstein}
\begin{equation}
    S=\int d^{N+1}x\sqrt{-g}\{R-\frac{4}{N-1}(\nabla\phi)^2-e^{-4a\phi/(N-1)}F^2-\mathcal{V}(\phi)\}, \label{action}
\end{equation}
where $R$ is the Ricci scalar, $F^2=F_{\mu\nu}F^{\mu\nu}$ is the Maxwell invariant, and $F_{\mu\nu}=\partial_{\mu}\mathcal{A}_{\nu}-\partial_{\nu}\mathcal{A}_{\mu}$ is the electromagnetic field strength tensor. We take the scalar potential to be
\begin{equation}
    \mathcal{V}(\phi)=e^{4b\phi/(N-1)}\Lambda,
\end{equation}
which is of Liouville type. In this action, $a$ and $b$ are the coupling constants that control the coupling of the dilaton field to the Maxwell field and to the $\Lambda$-dependent potential term, respectively. We note that in the non-dilatonic limit, the constant $\Lambda$ corresponds to the cosmological constant parameter. The choice $a=1$ is commonly associated with the dilaton-Maxwell coupling appearing in the low-energy effective action of heterotic string theory \cite{poletti1994global,rocha2018self}. However, in the present work we keep both coupling constants $a$ and $b$ arbitrary in order to study a more general Einstein-Maxwell-dilaton theory.

By varying the action (\ref{action}) with respect to the metric tensor $g_{\mu\nu}$, we obtain the Einstein field equations in $N+1$ dimensions \cite{ghezelbash2015cosmological}
\begin{eqnarray}
    \mathcal{G}_{\mu \nu} &\equiv& R_{\mu \nu}-\frac{1}{2}g_{\mu \nu} R-\frac{4}{N-1}[\nabla_\mu\phi\nabla_\nu\phi-\frac{1}{2}g_{\mu \nu}(\nabla\phi)^2]-e^{\frac{-4a\phi}{N-1}}[2F_{\mu \rho} F_\nu {}^\rho-\frac{1}{2}g_{\mu \nu} F^2]\nonumber\\
    &+& \, \frac{1}{2}e^{\frac{4b\phi}{N-1}}g_{\mu \nu}\Lambda=0.\label{einstein}
\end{eqnarray}

Varying the action (\ref{action}) with respect to the electromagnetic gauge field $\mathcal{A}_{\mu}$ and the dilaton field $\phi$ gives the Maxwell and dilaton field equations, respectively
\begin{equation}
    \mathcal{M}^\nu\equiv \nabla_\mu(e^{-4a\phi/(N-1)}F^{\mu \nu})=0, \label{maxwell}
\end{equation}
\begin{equation}
    {\cal D}\equiv \nabla^2\phi-\frac{b}{2}e^{4b\phi/(N-1)}\Lambda +\frac{a}{2}e^{-4a\phi/(N-1)}F^2=0. \label{dilaton}
\end{equation}

We consider a time-dependent spacetime ansatz in $N+1$ dimensions, with $N>2$, and with vanishing mixed time-space components $g_{ti}=0$, given by
\begin{equation}
    ds^2_{N+1}=-\frac{1}{H(t,\mathbf{x})^2}dt^2+H(t,\mathbf{x})^{\frac{2}{N-2}} R(t)^2\gamma_{ij}dx^idx^j, \label{linelement}
\end{equation}
where $\mathbf{x}$ denotes the coordinates on the $N$-dimensional spatial manifold described by the metric $\gamma_{ij}(\mathbf{x})$, with no assumption that these coordinates are Cartesian. We keep the spatial metric $\gamma_{ij}$ completely general at this stage. Moreover, $R(t)$ is an arbitrary time-dependent scale factor and $H(t,\mathbf{x})$ is an arbitrary metric function that depends on time and on the spatial coordinate $\mathbf{x}$. Any restrictions on $\gamma_{ij}(\mathbf{x})$, $H(t,\mathbf{x})$, and $R(t)$ will follow from the field equations. 

Throughout this work, the Greek indices $\mu,\nu,...$ denote components on the full $(N+1)$-dimensional spacetime, while Latin indices $i, j, ...$ denote components on the $N$-dimensional spatial manifold described by $\gamma_{ij}$. For the electromagnetic part, we consider a purely electric ansatz by assuming that the only nonzero component of the gauge potential $\mathcal{A}_{\mu}$ is 
\begin{equation}
    \mathcal{A}_t=\alpha R(t)^XH(t,\mathbf{x})^Y, \label{maxwellgauge}
\end{equation}
where $\alpha$, $X$, and $Y$ are constants. This choice allows the electric field to depend on the same geometric functions that appear in the metric ansatz. Finally, we take the dilaton field to be
\begin{equation}
    \phi(t,\mathbf{x})=-\frac{N-1}{4a}\ln{(H(t,\mathbf{x})^UR(t)^V)}, \label{dilatongauge}
\end{equation}
where $U$ and $V$ are constants. This logarithmic form of the dilaton is chosen to match the exponential couplings that appear in the action of the theory (\ref{action}). 

With the geometric, electromagnetic, and scalar field ansatz specified, we substitute them into the field equations (\ref{einstein})-(\ref{dilaton}) to derive the consistency conditions imposed by the Einstein, Maxwell, and dilaton equations. This will determine the allowed values of the constants $X$, $Y$, $U$, $V$, and $\alpha$, as well as restrictions on the metric functions $H(t,\mathbf{x})$ and $R(t)$, and the spatial metric $\gamma_{ij}$. 

For notational convenience, we introduce the following functions
\begin{align}
     A(t)\equiv\ln{R(t)}, \qquad  B(t,\mathbf{x})\equiv\ln{H(t,\mathbf{x})}.
\end{align}
This notation allows the derivatives of the metric functions to be written in terms of $A(t)$ and $B(t,\mathbf{x})$, and will be used throughout the following derivation.  With this notation, the electromagnetic gauge potential and the dilaton field take the following forms
\begin{equation}
    \mathcal{A}_t=\alpha e^{XA(t)+YB(t,\mathbf{x})}, \label{mgauge}
\end{equation}
\begin{equation}
     \phi(t,\mathbf{x})=-\frac{N-1}{4a}[UB(t,\mathbf{x})+V A(t)], \label{dgauge}
\end{equation}
respectively.

\section{General Dynamical Solutions in Einstein-Maxwell-dilaton Theory} \label{sec3}

In this section, we consider the most general branch within our ansatzes, in which the metric function depends on both time and the spatial coordinates $H=H(t,\mathbf{x})$. We show that the consistency of the Einstein-Maxwell-dilaton field equations fixes the constants appearing in the ansatzes and forces the coupling constants to be equal $a=b$. The remaining equations then reduce to master equations for the temporal and spatial functions, and constrain the base geometry. We then present the final form of the solution and discuss its conformal structure, as well as two important limiting cases of the theory.

\subsection{Reduction of the Field Equations and Consistency Conditions}

We begin with the Maxwell equation, which provides the first constraints on the electromagnetic ansatz. Substituting the gauge potential (\ref{mgauge}) and the dilaton ansatz (\ref{dgauge}) into the Maxwell equation (\ref{maxwell}), and lowering the index with the metric, we find that the independent nontrivial components are the spatial components $\mathcal{M}_i=0$ and the temporal component $\mathcal{M}_t=0$.

 The spatial components $\mathcal{M}_i=g_{ij}\mathcal{M}^j$, up to a nonvanishing overall factor, reduce to
\begin{equation}
   \partial_i\dot{B}+[(X+V+N-2)\dot{A}+(U+Y+2)\dot{B}]\partial_iB=0, \label{maxi}
\end{equation}
where $\dot{A}\equiv \frac{d}{dt}A(t)$, $\dot{B}\equiv\frac{\partial}{\partial t}B(t,\mathbf{x})$, and $\partial_i$ denotes differentiation with respect to the spatial coordinates. The temporal component of the Maxwell equation $\mathcal{M}_t=g_{tt}\mathcal{M}^t=0$ reduces to
\begin{equation}
    \Delta_{\gamma}B+(U+Y+2)(\nabla_{\gamma}B)^2=0, \label{Mtemp}
\end{equation}
where $\Delta_{\gamma}B\equiv \nabla^{(\gamma)}_i\nabla^{(\gamma)i}B=\frac{1}{\sqrt{\gamma}}\partial_i(\sqrt{\gamma}\gamma^{ij}\partial_jB)$ and $(\nabla_{\gamma}B)^2\equiv \gamma^{ij}\partial_iB\partial_jB$. Here $\nabla^{(\gamma)}_i$ is the covariant derivative associated with $\gamma_{ij}$, and $\gamma\equiv$ det$(\gamma_{ij})$.

We next consider the dilaton field equation, which contains contributions from both the Maxwell field and the Liouville potential and therefore provides additional conditions on the ansatzes. Using the dilaton ansatz (\ref{dgauge}) and the electromagnetic field strength, the dilaton field equation (\ref{dilaton}) reduces to
\begin{eqnarray}
    \mathcal{D} &=& \left(\frac{N-1}{4a}\right)e^{2B}\mathcal{T} -\frac{N-1}{4a}e^{-2A-\frac{2}{N-2}B}U\Delta_{\gamma}B-\frac{b}{2}\Lambda e^{-\frac{b}{a}(VA+UB)}
    \nonumber \\
    &-& a\alpha^2Y^2(\nabla_{\gamma}B)^2e^{(V+2X-2)A+(U+2Y+2-\frac{2}{N-2})B}=0, \label{deom}
\end{eqnarray}
where 
\begin{equation*}
    \mathcal{T}\equiv U\ddot{B}+V\ddot{A}+\Bigl(N\dot{A}+2\frac{N-1}{N-2}\dot{B}\Bigl)(V\dot{A}+U\dot{B}).
\end{equation*}
Using the temporal Maxwell equation (\ref{Mtemp}) to eliminate $\Delta_{\gamma}B$ from the dilaton equation (\ref{deom}), we get
\begin{eqnarray}
    \mathcal{D} &=& \left(\frac{N-1}{4a}\right)e^{2B}\mathcal{T} +\frac{N-1}{4a}e^{-2A-\frac{2}{N-2}B}U(U+Y+2)(\nabla_{\gamma}B)^2-\frac{b}{2}\Lambda e^{-\frac{b}{a}(VA+UB)}
    \nonumber \\
    &-& a\alpha^2Y^2(\nabla_{\gamma}B)^2e^{(V+2X-2)A+(U+2Y+2-\frac{2}{N-2})B}=0. \label{dilreduced}
\end{eqnarray}

The reduced dilaton field equation (\ref{dilreduced}) is now written in terms of the time derivative sector $\mathcal{T}$, the Liouville potential term, and two spatial gradient terms proportional to $(\nabla_{\gamma}B)^2$. For a nontrivial metric function $B(t,\mathbf{x})$, consistency of the ansatz requires these two gradient terms to have the same exponential dependence. Equating the corresponding exponents gives
\begin{align}
     V+2X=0, \qquad  U+2Y+2=0. \label{r1}
\end{align}
Equating the remaining coefficients of  $(\nabla_{\gamma}B)^2$ then fixes the electromagnetic gauge field parameter $\alpha$ in (\ref{mgauge}), as
\begin{equation}
    \alpha^2=-\frac{(N-1)U}{4a^2Y}. \label{r3}
\end{equation}
The remaining dilaton equation then becomes
\begin{equation}
     \mathcal{D}=\left(\frac{N-1}{4a}\right)e^{2B}\mathcal{T}-\frac{b}{2}\Lambda e^{-\frac{b}{a}(VA+UB)}=0. \label{dilremain} 
\end{equation}

Although the Maxwell and dilaton field equations provide important constraints on the ansatz parameters, the remaining constants and the conditions on the metric functions $H(t,\mathbf{x})$, $R(t)$, and the spatial metric $\gamma_{ij}$ are determined by the remaining field equations. Therefore, we now turn to the Einstein field equations given in (\ref{einstein}). For the metric ansatz (\ref{linelement}), the independent Einstein equations are the mixed $ti$, temporal $tt$, and spatial $ij$ components, given respectively by $\mathcal{G}_{ti}=0$, $\mathcal{G}_{tt}=0$, and $\mathcal{G}_{ij}=0$.

Starting with the mixed component, since $g_{ti}=0$, the terms proportional to the metric do not contribute. Moreover, for the electromagnetic gauge potential (\ref{mgauge}), the Maxwell contribution $F_{t\sigma} F_i{}^{\sigma}$ vanishes identically. Calculating the remaining Ricci and dilaton terms, the mixed Einstein equation becomes
\begin{equation}
    \mathcal{G}_{ti}=\partial_i\dot{B}+\Bigl[(N-2+\frac{N-2}{4a^2}UV)\dot{A}+(1+\frac{N-2}{4a^2}U^2)\dot{B} \Bigl]\partial_iB=0. \label{einti}
\end{equation}
This equation has the same differential structure in $B(t,\mathbf{x})$ as the spatial component of the Maxwell equation given in (\ref{maxi}). Comparing the corresponding coefficients and using the constraints obtained from the  dilaton equation in (\ref{r1}) and (\ref{r3}), we get
\begin{align}
    U=\frac{2a^2}{N-2}, \qquad Y=-1-\frac{a^2}{N-2}, \qquad \alpha^2=\frac{N-1}{2(a^2+N-2)}.
\end{align}
We note that these constants are not introduced by hand, but are fixed by the combined constraints of the dilaton field equation and the compatibility of the field equations $\mathcal{M}_i=0$ and $\mathcal{G}_{ti}=0$. 

Another important consequence of the Maxwell equations is the functional form of the metric function $B(t,\mathbf{x})$. Using the spatial component of the Maxwell (\ref{maxi}) and substituting the values of the constants fixed above, this equation can be written in the total-derivative form
\begin{equation}
    \partial_i\Bigl[e^{CB}(\dot{B}+\frac{D}{C}\dot{A})\Bigl]=0,
\end{equation}
where $C=1+\frac{a^2}{N-2}$ and $D=N-2+\frac{V}{2}$. This equation indicates that the quantity inside the bracket must be independent of the spatial coordinates and therefore, can be written as a function of time, $e^{CB}(\dot{B}+
    \frac{D}{C}\dot{A})=F(t)$. Equivalently, we can rewrite this as
\begin{equation}
    \partial_t(e^{CB+DA})=CF(t)e^{DA}.
\end{equation}
Integration with respect to time gives $e^{CB+DA}=h(\mathbf{x})+S(t)$, where $h(\mathbf{x})$ is the spatial integration function and $S(t)$ absorbs the time integral, $S(t)=C\int F(t)e^{DA}dt$.

Therefore, the metric function $B(t,\mathbf{x})$ is restricted to the form
\begin{equation}
    B(t,\mathbf{x})=\frac{1}{C}\ln{[h(\mathbf{x})+S(t)]}-\frac{D}{C}A(t). \label{Bform}
\end{equation}
Moreover, substituting this result for $B(t,\mathbf{x})$ into the temporal component of the Maxwell equation $\mathcal{M}_t=0$ in (\ref{Mtemp}), determines the spatial function $h(\mathbf{x})$ through the harmonic condition
\begin{equation}
    \Delta_{\gamma}h(\mathbf{x})=0.
\end{equation}
These results are essential for reducing the remaining Einstein equations, since they constrain the mixed $t$ and $\mathbf{x}$ dependence of $B(t,\mathbf{x})$ to the single combination $h(\mathbf{x})+S(t)$. 

We now turn to the remaining components of the Einstein field equations given in (\ref{einstein}), and evaluate the temporal component $\mathcal{G}_{tt}=0$  and the spatial component  $\mathcal{G}_{ij}=0$
\begin{eqnarray}
    \mathcal{G}_{tt} &=& \frac{1}{2}R(\gamma)e^{-2A-\frac{2(N-1)}{N-2}B}-\frac{1}{2}\Lambda e^{-\frac{b}{a}(VA+\frac{2a^2}{N-2}B)-2B}
    \nonumber \\
    &+& \frac{N-1}{2}\Bigl[(N-\frac{V^2}{4a^2})\dot{A}^2+\frac{N-a^2}{(N-2)^2}\dot{B}^2+\frac{2N-V}{N-2}\dot{A}\dot{B} \Bigl]=0, \label{eintt}
\end{eqnarray}
\begin{eqnarray}
    \mathcal{G}_{ij} &=& R_{ij}(\gamma)-\frac{1}{2}\gamma_{ij}R(\gamma)+\frac{1}{2}\gamma_{ij}\Lambda e^{(2A+\frac{2}{N-2}B)-\frac{b}{a}(VA+\frac{2a^2}{N-2}B)}-(N-1)e^{2A+\frac{2(N-1)}{N-2}B}\gamma_{ij}
    \nonumber \\
    &\times& \Bigl[\ddot{A}+\frac{\ddot{B}}{N-2} +\frac{N}{2}\dot{A}^2+\frac{2(N-1)}{N-2}\dot{A}\dot{B}+\frac{3N-4}{2(N-2)^2}\dot{B}^2 +\frac{\bigl(V\dot{A}+U\dot{B} \bigl)^2}{8a^2} \Bigl]=0, \label{einij}
\end{eqnarray}
where $R_{ij}(\gamma)$ and $R(\gamma)$ are the Ricci tensor and the Ricci scalar associated with the spatial base metric $\gamma_{ij}$, respectively.

The cosmological constant term appears algebraically in both the $tt$ and $ij$ components of the Einstein equations. We therefore solve the temporal equation $\mathcal{G}_{tt}=0$ for the $\Lambda$ dependent term. Substituting this result into the $ij$ component $\mathcal{G}_{ij}=0$ and using the Maxwell integrated form of the metric function $B(t,\mathbf{x})$ given in (\ref{Bform}), we get
\begin{eqnarray}
      \mathcal{G}_{ij} &=& R_{ij}(\gamma)-(N-1)e^{[2-\frac{2(N-1)D}{a^2+N-2}]A}\gamma_{ij}\Big\{\frac{2a^2-V}{2(a^2+N-2)}\bigl(\ddot{A}-\frac{V(N-2)}{2a^2}\dot{A}^2\bigl) \bigl[h(x)+S(t) \bigl]^{\frac{2(N-1)}{a^2+N-2}}  
      \nonumber \\
      &+& \frac{1}{a^2+N-2}\bigl(\ddot{S}-(N-2)\dot{A}\dot{S} \bigl)\bigl[h(x)+S(t) \bigl]^{\frac{2(N-1)}{a^2+N-2}-1} \Bigl\}=0.
\end{eqnarray}
  
The reduced $\mathcal{G}_{ij}$ equation now contains two distinct powers of the mixed function $h(\mathbf{x})+S(t)$. Since $R_{ij}(\gamma)$ depends only on the spatial base metric $\gamma_{ij}$, it cannot absorb the time dependence carried by the distinct powers of $h(\mathbf{x})+S(t)$. A cancellation between terms with different powers of $h(\mathbf{x})+S(t)$ would require an additional relation between $A(t)$, $S(t)$, and $h(\mathbf{x})$, corresponding to a separate special branch. Therefore, in the generic nontrivial branch considered here, the coefficient of the first independent contribution must vanish, which fixes $ V=2a^2$. Then, using the dilaton constraint $V+2X=0$ from (\ref{r1}), we also get
\begin{align}
    V=2a^2, \qquad  X=-a^2.
\end{align}

At this point, all constants appearing in the dilaton field (\ref{dilatongauge}) and electromagnetic gauge potential ansatzes (\ref{maxwellgauge}) have been fixed by the field equations.
With $V=2a^2$, the reduced spatial component of the Einstein equation becomes
\begin{equation}
    \mathcal{G}_{ij}=R_{ij}(\gamma)-(N-1)e^{-2(N-2)A}\gamma_{ij}\Bigl[\frac{1}{a^2+N-2}\bigl(\ddot{S}-(N-2)\dot{A}\dot{S}\bigl) \Bigl][h(\mathbf{x})+S(t)]^{\frac{N-a^2}{a^2+N-2}}=0. \label{redein}
\end{equation}
Taking the trace of this equation (\ref{redein}) with respect to $\gamma^{ij}$, we obtain
\begin{equation}
    R(\gamma)=\frac{N(N-1)}{a^2+N-2}e^{-2(N-2)A}\bigl(\ddot{S}-(N-2)\dot{A}\dot{S}\bigl)[h(\mathbf{x})+S(t)]^{\frac{N-a^2}{a^2+N-2}}. \label{einred2}
\end{equation}
Since $R(\gamma)$ depends only on the spatial base metric $\gamma_{ij}$ while the right-hand side generally depends on both $t$ and $\mathbf{x}$, consistency of the generic nontrivial branch requires
\begin{equation}
    \ddot{S}-(N-2)\dot{A}\dot{S}=0.
\end{equation}

This relation is an important temporal consistency condition and forms one of the master equations of the solution. With this condition imposed, the trace equation gives 
\begin{equation}
    R(\gamma)=0.
\end{equation}
Substituting $\ddot{S}-(N-2)\dot{A}\dot{S}=0$ back into the reduced spatial equation (\ref{redein}), we finally get
\begin{equation}
     R_{ij}(\gamma)=0.
\end{equation}
Therefore, the base metric must be Ricci flat. We note that this condition is not imposed as an assumption, but follows directly from the consistency of the reduced spatial Einstein equation. Consequently, the spatial sector of the solution is determined by a Ricci flat base metric $\gamma_{ij}$ and a harmonic function $h(\mathbf{x})$ on this base space.

Having fixed the constants in the ansatzes and obtained the constraints from the spatial Einstein equation, the only remaining field equations are the reduced dilaton equation (\ref{dilremain}) and the temporal component of the Einstein equation $\mathcal{G}_{tt}$ (\ref{eintt}). We first substitute the Maxwell integrated form of the metric function $B(t,\mathbf{x})$ and the constants found above into the reduced dilaton equation (\ref{dilremain}). This gives
\begin{equation}
    \mathcal{D}=\Lambda e^{2(N-2)A}\bigl(h(\mathbf{x})+S(t)\big)^{\frac{-2(N-2+ab)}{a^2+N-2}}-\frac{a(N-1)(N-a^2)}{b(a^2+N-2)^2}\frac{\dot{S}^2}{(h(\mathbf{x})+S(t))^2}=0. \label{dillast}
\end{equation}
Applying the same substitution to the $tt$ component of the Einstein equation (\ref{eintt}) gives
\begin{equation}
    \mathcal{G}_{tt}=\Lambda e^{2(N-2)A}\bigl(h(\mathbf{x})+S(t)\big)^{\frac{-2(N-2+ab)}{a^2+N-2}}-\frac{(N-1)(N-a^2)}{(a^2+N-2)^2}\frac{\dot{S}^2}{(h(\mathbf{x})+S(t))^2}=0. \label{einlast}
\end{equation}
We realize that both the dilaton (\ref{dillast}) and the temporal component of the Einstein equations (\ref{einlast}) have the same $\Lambda$ dependent structure and the same dependence on $A(t)$ and $h(\mathbf{x})+S(t)$. The only difference is the overall factor $a/b$ multiplying  the second term in the dilaton equation (\ref{dillast}). Therefore, for the generic nontrivial branch, these two equations are compatible only when
\begin{equation}
    a=b.
\end{equation}

This is an important consequence of the ansatz structure considered in this section. The condition $a=b$ indicates that the dilaton couples to the Maxwell field and to the cosmological constant $\Lambda$ with the same coupling strength. Although we started with the Einstein-Maxwell-dilaton action with two independent coupling constants $a$ and $b$, the field equations require the coupling constants to coincide within this branch. 

\subsection{Final Form of the Solution and Conformal Structure}

Having established that the consistency of the field equations requires the equal coupling constants $a=b$, we now collect the results. The constants appearing in the dilaton and Maxwell ansatzes are determined as
\begin{align*}
    U &= \frac{2a^2}{N-2},  \qquad V=2a^2,
\end{align*}
and
\begin{align*}
     & X = -a^2, \qquad  Y=-1-\frac{a^2}{N-2},  \qquad \alpha^2=\frac{N-1}{2(a^2+N-2)}.
\end{align*}
The Maxwell equation (\ref{maxi}) also determines the separable structure of the metric function $B(t,\mathbf{x})$ as
\begin{equation*}
    B(t,\mathbf{x})=\frac{N-2}{a^2+N-2}\ln{\bigl(h(\mathbf{x})+S(t)\bigl)}-(N-2)A(t).
\end{equation*}
Equivalently, this relation becomes
\begin{equation}
    H(t,\mathbf{x})=R(t)^{-(N-2)}[h(\mathbf{x})+S(t)]^{\frac{N-2}{a^2+N-2}},
\end{equation}
where we used $H(t,\mathbf{x})=e^{B(t,\mathbf{x})}$ and $R(t)=e^{A(t)}$. Therefore, the original metric functions $H(t,\mathbf{x})$ and $R(t)$ are not independent in this branch. Instead, their dependence is controlled by a single combination 
\begin{equation}
    \Psi(t,\mathbf{x})=h(\mathbf{x})+S(t).
\end{equation}

The remaining field equations require the spatial function $h(\mathbf{x})$ to be harmonic with respect to the base metric $\gamma_{ij}$, while the base metric itself must be Ricci flat
\begin{equation*}
    \Delta_{\gamma}h=0,  \qquad R_{ij}(\gamma)=0.
\end{equation*}
Moreover, the temporal sector is governed by
\begin{equation*}
    \ddot{S}-(N-2)\dot{A}\dot{S}=0.
\end{equation*}
Integrating this relation, we obtain $\dot{S}=\eta e^{(N-2)A}=\eta R(t)^{N-2}$, where $\eta$ is an integration constant. Using this result and imposing the condition $a=b$, the reduced dilaton equation (\ref{dillast}) and the temporal Einstein field equation (\ref{einlast}) give the same remaining condition
\begin{equation}
    \Lambda e^{2(N-2)A}=\frac{(N-1)(N-a^2)}{(a^2+N-2)^2}\dot{S}^2. \label{lamS}
\end{equation}
Substituting the integrated expression for $\dot{S}$, this condition fixes the cosmological constant in terms of the integration constant $\eta$ as
\begin{equation}
    \Lambda=\eta^2 \frac{(N-1)(N-a^2)}{(a^2+N-2)^2}. \label{cosmo}
\end{equation}
This expression (\ref{cosmo}) shows that the sign of the cosmological constant is controlled by the value of the coupling constant $a$.

In terms of $\Psi$, the line element (\ref{linelement}) can be written as
\begin{equation}
    ds^2=-R(t)^{2(N-2)}\Psi^{-\frac{2(N-2)}{a^2+N-2}}dt^2+\Psi^{\frac{2}{a^2+N-2}}\gamma_{ij}dx^idx^j,
    \end{equation}
which already indicates an important geometric feature of the solution. The spatial metric is conformal to the Ricci-flat base metric $\gamma_{ij}$
\begin{equation}
    g_{ij}=\Psi^{\frac{2}{a^2+N-2}}\gamma_{ij}.
\end{equation}
Writing the temporal relation $\dot{S}=\eta R(t)^{N-2}$ in differential form as $ dS=\eta R^{N-2}dt$ shows that the metric function $R(t)$ can be absorbed into a redefined time coordinate. Therefore, the conformal structure becomes more explicit by introducing a new time coordinate $\tau$, defined by
\begin{equation}
    d\tau=R(t)^{N-2}dt. 
\end{equation}
In this parametrization, the temporal equation gives $S(\tau)=\eta \tau +S_0$, where $S_0$ is an integration constant. The conformal potential $\Psi$ then becomes
\begin{equation}
    \Psi(\tau,\mathbf{x})=h(\mathbf{x})+\eta\tau+S_0,
\end{equation}
and the metric reduces to
\begin{equation}
    ds^2=-\Psi^{-\frac{2(N-2)}{a^2+N-2}}d\tau^2+\Psi^{\frac{2}{a^2+N-2}}\gamma_{ij}dx^idx^j. \label{conformal}
\end{equation}

This is the conformal potential form of the solution. It shows that the original two function structure of the metric described by $H(t,\mathbf{x})$ and $R(t)$ is reduced by the field equations to the single combination $\Psi=h+\eta\tau+S_0$. After the time reparameterization, the metric function $R(t)$ no longer appears explicitly, while the spatial geometry remains conformal to the Ricci flat base metric $\gamma_{ij}$. Therefore, once the Ricci flat base metric and the harmonic spatial function $h(\mathbf{x})$ are specified, the spacetime geometry is determined by the single composite function $\Psi$, which fixes the conformal scaling of the spatial metric.

A possible singular behaviour of the geometry is also naturally encoded in this potential. In particular, the hypersurface $\Psi(\tau,\mathbf{x})=0$ leads to a degenerate behaviour in the metric. This can be seen more explicitly from the Ricci scalar, which is given by 
\begin{equation}
    \mathcal{R}=\frac{N\eta^2(N+1-2a^2)}{(a^2+N-2)^2}\Psi^{\frac{-2a^2}{a^2+N-2}}+\frac{2a^2-(N-2)(N-3)}{(a^2+N-2)^2}(\nabla_{\gamma}h)^2\Psi^{-\frac{2(a^2+N-1)}{a^2+N-2}}.
\end{equation}

The Ricci scalar shows that the curvature is controlled by the conformal potential $\Psi(\tau,\mathbf{x})$. Therefore, for generic values of the coupling constant and for a nontrivial harmonic function $h(\mathbf{x})$, the Ricci scalar diverges as $\Psi\to 0$. The hypersurface $\Psi=0$ then corresponds to a curvature singularity. Moreover, in the asymptotic region $\Psi\to \infty$, the Ricci scalar approaches zero. We note that special values of the coupling constant or special choices of $h(\mathbf{x})$ may remove some contributions to $\mathcal{R}$, but this does not necessarily remove the singularity from the full geometry. 

To confirm this behaviour, we also calculate the Kretschmann invariant. Since its expression is lengthy, we do not present it explicitly here. We only note that it contains several terms with negative powers of $\Psi$, which are generically nonzero. Therefore, the Kretschmann invariant also diverges as $\Psi\to 0$, further confirming the singular behaviour of the geometry.

We summarize the results of this section in the following proposition:

\textbf{Proposition}: Within the ansatz class considered above, the Einstein-Maxwell-dilaton field equations admit a generic mixed time-space branch only in the equal coupling sector $a=b$. In this branch, the field equations reduce to the master system derived above: the base metric $\gamma_{ij}$ is Ricci flat, $h(\mathbf{x})$ is harmonic on the base space, and the temporal function satisfies $dS=\eta R(t)^{N-2}dt$, with $\eta$ related to the cosmological constant by (\ref{cosmo}).

Equivalently, after introducing $d\tau=R(t)^{N-2}dt$, the metric can be written in the conformal potential form (\ref{conformal}), with $\Psi(\tau,\mathbf{x})=h(\mathbf{x})+\eta\tau+S_0$. In this form, the original two-function structure of the metric is reduced to a single potential $\Psi$, which controls the spatial conformal factor and the temporal part of the metric through different powers.  

We emphasize that the conditions summarized above are not imposed a priori. They arise from the consistency of the Maxwell, dilaton, and Einstein field equations within the ansatz class considered in this section.

\subsection{Einstein-Maxwell Limit}

In this subsection, we consider the Einstein-Maxwell theory with a cosmological constant $\Lambda$ in $N+1$ dimensions. Since the limit $a=b=0$ makes the dilaton ansatz (\ref{dilatongauge}) diverge, we first remove the dilaton field by taking $\phi(t,\mathbf{x})=0$. The remaining field equations then describe the non-dilatonic Einstein-Maxwell theory in the vanishing coupling limit $a=b\rightarrow 0$.

The action obtained from the Einstein-Maxwell-dilaton theory in this limit is given by
\begin{equation}
    S=\int d^{N+1}x\sqrt{-g}\{R-F^2-\Lambda \}.
\end{equation}
For this theory, we consider the same metric ansatz as in (\ref{linelement})
\begin{equation*}
    ds^2_{N+1}=-\frac{1}{H(t,\mathbf{x})^2}dt^2+H(t,\mathbf{x})^{\frac{2}{N-2}} R(t)^2\gamma_{ij}dx^idx^j.
\end{equation*}
Using the results obtained above and taking the non-dilatonic limit $a=b\rightarrow 0$, the constants appearing in the electromagnetic ansatz (\ref{maxwellgauge}) reduce to
\begin{align}
    X=0, \qquad Y=-1, \qquad \alpha^2=\frac{N-1}{2(N-2)}.
\end{align}
Therefore, the electromagnetic gauge ansatz (\ref{maxwellgauge}) becomes
\begin{equation}
     \mathcal{A}_t=\frac{\alpha}{H(t,\mathbf{x})}.
\end{equation}

The field equations for the functions $h(\mathbf{x})$ and $S(t)$ keep the same form as before
\begin{align*}
    \Delta_{\gamma}h(\mathbf{x})=0, \qquad \dot{S}=\eta R^{N-2},
\end{align*}
where in the limit $a=b\rightarrow0$, equation (\ref{cosmo}) fixes the cosmological constant in terms of the integration constant $\eta$ as 
\begin{equation}
    \Lambda=\eta^2 \frac{N(N-1)}{(N-2)^2}.
\end{equation}
Moreover, the metric function $H(t,\mathbf{x})$ now takes the following form
\begin{equation}
    H(t,\mathbf{x})=R(t)^{-(N-2)}\bigl [h(\mathbf{x})+S(t)\bigl].
\end{equation}
We note that the base geometry $\gamma_{ij}$ remains Ricci flat, which follows from the consistency of the reduced field equations.

\subsection{Einstein-Maxwell-dilaton with vanishing Cosmological Constant $\Lambda=0$}

In this subsection, we briefly discuss the Einstein-Maxwell-dilaton theory in the absence of the Liouville potential. For this purpose, we consider the limit in which the cosmological constant vanishes $\Lambda=0$. The action of the theory in this limit becomes
\begin{equation}
    S=\int d^{N+1}x\sqrt{-g}\{R-\frac{4}{N-1}(\nabla\phi)^2-e^{-4a\phi/(N-1)}F^2\}.
\end{equation}

We now impose this limit on the results derived for the general branch of the metric function $B=B(t,\mathbf{x})$. Since the cosmological constant appears only in the remaining $\mathcal{G}_{tt}$, $\mathcal{G}_{ij}$, and the dilaton equation, the previous results leading to the form of $B(t,\mathbf{x})$ remain unchanged. In the limit $\Lambda=0$, the remaining temporal equation given in (\ref{lamS}) becomes
\begin{equation}
    \frac{(N-1)(N-a^2)}{(a^2+N-2)^2}\dot{S}^2=0. \label{nolam}
\end{equation}

For the generic case $a^2\neq N$, this equation requires $\dot{S}=0$. Therefore, the temporal function is forced to be a constant $ S(t)=S_0$. With this condition, the temporal equation obtained from the spatial Einstein field equation $ \ddot{S}-(N-2)\dot{A}\dot{S}=0$ is satisfied trivially. Therefore, the field equations do not further restrict the time dependent scale factor $R(t)$ in this limit. The metric function $B(t,\mathbf{x})$ then reduces to 
\begin{equation}
    B(t,\mathbf{x})=\frac{N-2}{a^2+N-2}\ln{\bigl(h(\mathbf{x})+S_0\bigl)}-(N-2)A(t).
\end{equation}
Equivalently, in terms of $H(t,\mathbf{x})=e^{B(t,\mathbf{x})}$, we get
\begin{equation}
    H(t,\mathbf{x})=R(t)^{-(N-2)}[h(\mathbf{x})+S_0]^{\frac{N-2}{a^2+N-2}}.
\end{equation}

Substituting this result into the original line element (\ref{linelement}), we find
\begin{equation}
    ds^2=-R(t)^{2(N-2)}[h(\mathbf{x})+S_0]^{\frac{-2(N-2)}{a^2+N-2}}dt^2+[h(\mathbf{x})+S_0]^{\frac{2}{a^2+N-2}}\gamma_{ij}dx^idx^j,
\end{equation}
indicating that the spatial part of the metric is now independent of time. By defining a new time coordinate $\tau$ as
\begin{equation}
    d\tau=R(t)^{N-2}dt,
\end{equation}
the line element takes the static form
\begin{equation}
    ds^2=-[h(\mathbf{x})+S_0]^{\frac{-2(N-2)}{a^2+N-2}}d\tau^2+[h(\mathbf{x})+S_0]^{\frac{2}{a^2+N-2}}\gamma_{ij}dx^idx^j.
\end{equation}

Therefore, in the vanishing Liouville potential limit with $a^2\neq N$, the function $S(t)$ is forced to be constant. The spatial geometry becomes independent of time, while the remaining arbitrary function $R(t)$ can be absorbed into a redefinition of the time coordinate. The line element in the $\Lambda=0$ limit of the mixed branch $B(t,\mathbf{x})$ is effectively static in the new time coordinate. We note that for the special value of the coupling constant $a^2=N$, equation (\ref{nolam}) is identically satisfied. In this case, $S(t)$ is no longer required to be constant. Instead, it is constrained by the remaining temporal equation $\ddot{S}-(N-2)\dot{A}\dot{S}=0$, which allows the solution to retain the same form as in the general dynamical branch.

The analysis of this section shows that the mixed time-space branch of the metric function $H(t,\mathbf{x})$ forces the equal coupling constants $a=b$ and reduces the geometry to the conformal potential $\Psi(\tau,\mathbf{x})=h(\mathbf{x})+\eta\tau+S_0$. In the next section, we consider a pure spatial metric function $H(\mathbf{x})$ to explore other possibilities.

\section{Purely Spatial Branch with General Coupling Constants} \label{sec4}

In this section, we continue with the same Einstein-Maxwell-dilaton theory defined by the action in equation (\ref{action}), with the field equations given in (\ref{einstein})-(\ref{dilaton}). The difference from the mixed time-space branch is that we now consider the metric function $H$ to be purely spatial $H=H(\mathbf{x})$. This choice allows us to examine whether the constraints found in the previous section persist, or whether the field equations lead to a different set of restrictions.

\subsection{Reduction of the Field Equations for the Purely Spatial Branch}

We begin by applying the purely spatial restriction to the line element given in (\ref{linelement}). The metric then becomes
\begin{equation}
    ds^2_{N+1}=-\frac{1}{H(\mathbf{x})^2}dt^2+H(\mathbf{x})^{\frac{2}{N-2}} R(t)^2\gamma_{ij}dx^idx^j, \label{linelement2}
\end{equation}
where $\mathbf{x}$ represents the coordinates on the $N$-dimensional spatial manifold described by a general base metric $\gamma_{ij}(\mathbf{x})$. The metric ansatz (\ref{linelement2}) is of a familiar form that has been widely used in the construction of exact solutions in gravitational theories, such as in cosmology, Einstein-Maxwell, and Einstein-Maxwell-dilaton theories \cite{harrison1968new, ida2007cosmological, butler2019minimal}. The time dependent scale factor $R(t)$ represents the overall time evolution of the spatial geometry, and the spatial function $H(\mathbf{x})$ describes the inhomogeneous deformation of the spacetime. 

Following the notations of the previous section, we define
\begin{align}
    A(t)\equiv \ln{R(t)}, \qquad  B(\mathbf{x})\equiv \ln{H(\mathbf{x})}.
\end{align}
With these definitions, the corresponding ansatzes for the electromagnetic potential and the dilaton field are taken to be
\begin{equation}
    \mathcal{A}_t=\alpha e^{XA(t)+YB(\mathbf{x})}, \label{mgauge2}
\end{equation}
\begin{equation}
     \phi(t,\mathbf{x})=-\frac{N-1}{4a}[UB(\mathbf{x})+V A(t)], \label{dgauge2}
\end{equation}
respectively.

The first constraints follow from the Maxwell field equations. Substituting the metric, electromagnetic gauge, and dilaton ansatzes into the Maxwell equation $\mathcal{M}^{\mu}=0$ given in (\ref{maxwell}), the temporal component becomes
\begin{equation}
    \mathcal{M}^t=\Delta_{\gamma}B+(U+Y+2)(\nabla_{\gamma}B)^2=0, \label{mt2}
\end{equation}
where $(\nabla_{\gamma}B)^2\equiv\gamma^{ij}\partial_iB\partial_jB$, and $\Delta_{\gamma}B\equiv \nabla^{(\gamma)}_i\nabla^{(\gamma)i}B$, with $\nabla^{(\gamma)}_i$ the covariant derivative associated with the base metric $\gamma_{ij}$. Moreover, since $B=B(\mathbf{x})$, the spatial component of the Maxwell equation $\mathcal{M}^i=0$ reduces to
\begin{equation}
    \mathcal{M}^i=(N+X+V-2)\dot{A}\partial^iB=0,
\end{equation}
which for nontrivial metric functions $A(t)$ and $B(\mathbf{x})$, gives
\begin{equation}
    N+X+V-2=0. \label{xv}
\end{equation}

The dilaton field equation provides the next set of consistency conditions. Evaluating the scalar and electromagnetic contributions using the ansatzes above, the dilaton equation (\ref{dilaton}) becomes
\begin{eqnarray}
    \mathcal{D} &=& \frac{N-1}{4a}Ve^{2B}\bigl(\ddot{A}+N\dot{A}^2\bigl) -\frac{N-1}{4a}Ue^{-2A-\frac{2}{N-2}B}\Delta_{\gamma}B-\frac{b}{2}\Lambda e^{-\frac{b}{a}(VA+UB)}
    \nonumber \\
    &-& a\alpha^2Y^2e^{(V+2X-2)A+(U+2Y+2-\frac{2}{N-2})B}(\nabla_{\gamma}B)^2=0. \label{dred2}
\end{eqnarray}
We use the temporal component of the Maxwell (\ref{mt2}) to eliminate $\Delta_{\gamma}B$ from the dilaton equation (\ref{dred2}). The remaining spatial-derivative terms then become proportional to $(\nabla_{\gamma}B)^2$. Since $A(t)$ and $B(\mathbf{x})$ are independent functions, the spatial-derivative terms cannot be canceled by the purely temporal contribution proportional to $(\ddot{A}+N\dot{A}^2)$ or by the Liouville potential term. Therefore, the consistency requires the two gradient contributions to have the same exponential dependence on $A(t)$ and $B(\mathbf{x})$, which gives 
\begin{align}
     V+2X=0,  \qquad U+2Y+2=0. \label{uy2}
\end{align}
With the constraint obtained from the spatial Maxwell equation $\mathcal{M}^i$ in (\ref{xv}), the first relation fixes
\begin{align}
     X=N-2, \qquad V=-2(N-2).
\end{align}

We note that these values differ from those we obtained in the previous section. This difference is a consequence of restricting the metric function to depend only on the spatial coordinates, $B=B(\mathbf{x})$. Substituting these relations back into the dilaton equation, the remaining gradient term must vanish independently. This gives
\begin{equation}
    \frac{N-1}{4a}U(U+Y+2)-a\alpha^2Y^2=0, \label{alph}
\end{equation}
and the dilaton equation then reduces to
\begin{equation}
    \mathcal{D} = -\frac{(N-2)(N-1)}{2a}e^{2B}(\ddot{A}+N\dot{A}^2)-\frac{b}{2}\Lambda e^{-\frac{b}{a}(-(N-2)A+UB)}=0. \label{d3}
\end{equation}

To determine the remaining constants appearing in the ansatzes and the corresponding constraints on the metric functions, we now turn to the Einstein field equations (\ref{einstein}). We begin with the mixed $ti$ component, $\mathcal{G}_{ti}=0$. Since $g_{ti}=0$ for the metric ansatz, and since the Maxwell contribution $F_{t\sigma} F_i{}^{\sigma}$ vanishes for the gauge ansatz (\ref{mgauge2}), the mixed Einstein equation reduces to 
\begin{equation}
    \mathcal{G}_{ti}=-(N-1)\Bigl[1-(N-2)\frac{U}{2a^2}\Bigl]\dot{A}\partial_iB=0.
\end{equation}
For nontrivial metric functions $A(t)$ and $B(\mathbf{x})$, this equation fixes the constant $U$. Combining this result with the algebraic relations obtained from the dilaton equation in (\ref{uy2}) and (\ref{alph}), the remaining constants are determined as
\begin{equation}
    U=\frac{2a^2}{N-2}, \qquad Y=-1-\frac{a^2}{N-2}, \qquad \alpha^2=\frac{N-1}{2(a^2+N-2)}.
\end{equation}

By substituting these results into (\ref{d3}), the remaining dilaton equation becomes
\begin{equation}
     \mathcal{D} = -\frac{(N-2)(N-1)}{2a}e^{2B}(\ddot{A}+N\dot{A}^2)-\frac{b}{2}\Lambda e^{-\frac{b}{a}(-2(N-2)A+\frac{2a^2}{N-2}B)}=0. \label{ddd}
\end{equation}
For a nontrivial metric function $B(\mathbf{x})$, the two terms in the dilaton equation cannot cancel unless they carry the same exponential dependence on $B(\mathbf{x})$. Comparing the coefficients of $B(\mathbf{x})$ in the exponential factors gives
\begin{equation}
    ab=-(N-2),
\end{equation}
which shows that the coupling constants $a$ and $b$ are not independent for this class of solutions. We note that in the previous section with $B=B(t,\mathbf{x})$, the field equations forced the coupling constants to be equal $a=b$. For the purely spatial metric function $B=B(\mathbf{x})$, the field equations instead lead to the constraint $ab=-(N-2)$, allowing the coupling constants to be more general while still being related.

Substituting the relation $ab=-(N-2)$ into the reduced dilaton equation (\ref{ddd}) and dropping the nonzero overall factor $\frac{(N-2)}{2a}e^{2B}$, we get
\begin{equation}
    -(N-1)(\ddot{A}+N\dot{A}^2)+\Lambda e^{-2\frac{(N-2)^2}{a^2}A}=0. \label{dlast}
\end{equation}

At this stage, the field equations have determined the constants $U$, $V$, $Y$, $X$, and $\alpha$ that appear in the ansatzes (\ref{mgauge2})-(\ref{dgauge2}). We now calculate the remaining field equations, namely temporal and spatial components of the Einstein field equations in order to find constraints on the metric functions $A(t)$ and $B(\mathbf{x})$, as well as the base geometry $\gamma(\mathbf{x})$.

\subsection{Master Equations and Geometric Structure}

Using the Einstein equation (\ref{einstein}) and evaluating the corresponding geometric and matter contributions, the spatial component $\mathcal{G}_{ij}=0$ becomes
\begin{equation}
    \mathcal{G}_{ij}=R_{ij}(\gamma)-\frac{1}{2}\gamma_{ij}R(\gamma)+e^{\frac{2(N-1)}{N-2}B}\gamma_{ij}G(t)e^{2A}=0, \label{einij2}
\end{equation}
where $R_{ij}(\gamma)$ and $R(\gamma)$ are the Ricci tensor and the Ricci scalar associated with the spatial base geometry $\gamma_{ij}$, respectively, and
\begin{equation}
    G(t)\equiv-(N-1)\Bigl[\ddot{A}+\frac{1}{2}(N+\frac{(N-2)^2}{a^2})\dot{A}^2 \Bigl]+\frac{1}{2}\Lambda e^{-\frac{2(N-2)^2}{a^2}A}.
\end{equation}

The $ij$ component of the Einstein equation (\ref{einij2}) has a separable structure. The first two terms depend only on the base metric $\gamma_{ij}$, while the last term contains the product of a spatial factor $e^{\frac{2(N-1)}{N-2}B}\gamma_{ij}$ and a purely temporal part $G(t)e^{2A}$. For nontrivial metric functions $A(t)$ and $B(\mathbf{x})$, the consistency of the equation requires the separated factors to be related by a separation constant $\lambda$ as
\begin{align}
    R_{ij}(\gamma)-\frac{1}{2}\gamma_{ij}R(\gamma)=\lambda e^{\frac{2(N-1)}{N-2}B}\gamma_{ij}, \qquad -G(t)e^{2A}=\lambda. \label{s2}
\end{align}
From the contracted Bianchi identity associated with the base metric $\gamma_{ij}$, and the metric compatibility of $\gamma_{ij}$, we have
\begin{align}
    \nabla_{\gamma}^i\Bigl(R_{ij}(\gamma)-\frac{1}{2}\gamma_{ij}R(\gamma)\Bigl)=0, \qquad \nabla_{\gamma}^i
(\gamma_{ij})=0.\end{align}
Applying $\nabla_{\gamma}^i$ to the first relation in (\ref{s2}), we get
\begin{equation}
    0=\lambda\frac{2(N-1)}{N-2}e^{\frac{2(N-1)}{N-2}B} \partial_jB.
\end{equation}

For a nontrivial function $B(\mathbf{x})$, this equation requires the separation constant to vanish $\lambda=0$. The first relation in (\ref{s2}) then reduces to $ R_{ij}(\gamma)-\frac{1}{2}\gamma_{ij}R(\gamma)=0$. Since $N>2$, taking the trace gives $R(\gamma)=0$, and hence 
\begin{equation}
   R_{ij}(\gamma)=0.
\end{equation}
Therefore, consistency of the field equations with the proposed ansatzes requires the base geometry $\gamma_{ij}$ to be Ricci flat. We note that this strong restriction was not imposed a priori, but is forced by the field equations.

Since $e^{2A}\neq0$, setting $\lambda=0$ in the second relation in equation (\ref{s2}) implies $G(t)=0$. Therefore, the temporal constraint becomes

\begin{equation}
    G(t)=-(N-1)\Bigl[\ddot{A}+\frac{1}{2}(N+\frac{(N-2)^2}{a^2})\dot{A}^2 \Bigl]+\frac{1}{2}\Lambda e^{-\frac{2(N-2)^2}{a^2}A}=0. \label{temij}
\end{equation}
We now consider the $tt$ component of the Einstein equation. Substituting the ansatzes and the constraints found above into (\ref{einstein}), we get
\begin{equation}
    \mathcal{G}_{tt}=(\frac{N-1}{2})(N-\frac{(N-2)^2}{a^2})\dot{A}^2+e^{-2A-\frac{2(N-1)}{N-2}B}Q(x)-\frac{1}{2}\Lambda e^{-\frac{2(N-2)^2}{a^2}A}=0, \label{eint2}
\end{equation}
where the purely spatial part of the equation is denoted by $Q(x)$
\begin{equation}
    Q(\mathbf{x})\equiv -\frac{N-1}{N-2}\Delta_{\gamma}B-(N-1)\frac{a^2+N-2}{(N-2)^2}(\nabla_{\gamma}B)^2.
\end{equation}

The temporal component of the Einstein equation (\ref{eint2}) also has a separable structure. We can rewrite it as
\begin{equation}
    e^{-\frac{2(N-1)}{N-2}B}Q(\mathbf{x})=-e^{2A}\Bigl[ (\frac{N-1}{2})(N-\frac{(N-2)^2}{a^2})\dot{A}^2-\frac{1}{2}\Lambda e^{-\frac{2(N-2)^2}{a^2}A}\Bigl].
\end{equation}
Since the left-hand side depends only on the spatial coordinates, and the right-hand side depends only on time, both sides of the equation must be equal to a separation constant, which we denote by $\sigma$. Therefore, we get
\begin{equation}
    Q(\mathbf{x})=\sigma e^{\frac{2(N-1)}{N-2}B}, \label{Q}
\end{equation}
and
\begin{equation}
     (\frac{N-1}{2})(N-\frac{(N-2)^2}{a^2})e^{2A}\dot{A}^2-\frac{1}{2}\Lambda e^{(2-\frac{2(N-2)^2}{a^2})A}=-\sigma. \label{temptt}
\end{equation}

We now use the remaining equations to determine the separation constant $\sigma$. For this purpose, we first isolate the $\Lambda$ dependent term from the reduced dilaton equation in (\ref{dlast}), which gives
\begin{equation}
    \Lambda e^{-2\frac{(N-2)^2}{a^2}A}=(N-1)(\ddot{A}+N\dot{A}^2). \label{lamdil}
\end{equation}
Using this relation to eliminate the $\Lambda$ dependent term in the temporal part of the Einstein equation $\mathcal{G}_{ij}=0$, given in (\ref{temij}), we obtain the master differential equation that determines the behaviour of the metric function $A(t)$, as
\begin{equation}
    \ddot{A}+\frac{(N-2)^2}{a^2}\dot{A}^2=0. \label{mastertem}
\end{equation}
This equation provides the main temporal constraint on the metric function $A(t)$. Its role in fixing the form of the solution will be discussed later in this section, after deriving the remaining constraints.

Using the expression for the $\Lambda$ dependent term given in (\ref{lamdil}), we eliminate this term from the separated temporal equation obtained from the $tt$ component of the Einstein equation $\mathcal{G}_{tt}=0$, given in (\ref{temptt}). This gives
\begin{equation}
   -\frac{(N-1)}{2} e^{2A}\Bigl[\ddot{A}+\frac{(N-2)^2}{a^2}\dot{A}^2 \Bigl]=-\sigma.
\end{equation}
The expression in the bracket is exactly the master equation obtained in (\ref{mastertem}). Therefore, consistency requires the separation constant to vanish $\sigma=0$.
Using this result for the separation constant $\sigma$, we return to the spatial equation obtained from the separated $\mathcal{G}_{tt}=0$, given in (\ref{Q}). Since $\sigma=0$, this equation reduces to $Q(\bm{x})=0$. Removing the overall nonzero factor, we get
\begin{equation}
    \Delta_{\gamma}B+\frac{a^2+N-2}{(N-2)}(\nabla_{\gamma}B)^2=0.
\end{equation}
This is exactly the same spatial constraint obtained from the temporal Maxwell equation $\mathcal{M}^t=0$, given in (\ref{mt2}).

\subsection{Summary and Interpretation of the Reduced System}

For the proposed ansatzes in the Einstein-Maxwell-dilaton theory in $N+1$ dimensions, we showed that the field equations fix the constants $U$ and $V$ appearing in the  dilaton ansatz (\ref{dgauge2}) as
\begin{align*}
    U=\frac{2a^2}{N-2}, \quad V=-2(N-2).
\end{align*}
Moreover, the field equations fix the constants $X$, $Y$, and $\alpha$ appearing in the electromagnetic gauge ansatz (\ref{mgauge2}) as
\begin{align*}
    X=N-2, \qquad Y=-1-\frac{a^2}{N-2}, \qquad \alpha^2=\frac{N-1}{a^2+N-2}.
\end{align*}
Another important result is the relation obtained between the coupling constants $a$ and $b$. We found that consistency of the field equations requires 
\begin{equation*}
    ab=-(N-2),
\end{equation*}
where $N$ is the number of spatial dimensions.

We also showed that the spatial base geometry $\gamma_{ij}$ must be Ricci flat. Therefore, the Ricci flatness of $\gamma_{ij}$ emerges as a consistency condition of the reduced field equations. The remaining components of the field equations further lead to the following spatial constraint on metric function $B(\mathbf{x})$
\begin{equation*}
    \Delta_{\gamma}B+\frac{a^2+N-2}{(N-2)}(\nabla_{\gamma}B)^2=0.
\end{equation*}
Using the relation $B(\mathbf{x})=\ln{H(\mathbf{x})}$, we can rewrite this differential equation as $ H\Delta_{\gamma}H+\frac{a^2}{N-2}(\nabla_{\gamma}H)^2=0$. Equivalently, this equation can be written as the harmonic condition

\begin{equation}
   \Delta_{\gamma}(H^{\frac{a^2+N-2}{N-2}})=0,
\end{equation}
which shows that the appropriate power of the metric function $H(\mathbf{x})$, rather than $H(\mathbf{x})$ itself, is harmonic on the Ricci flat base geometry $\gamma_{ij}$. We may therefore write
\begin{equation}
    H(\mathbf{x})^{\frac{a^2+N-2}{N-2}}=h(\mathbf{x})+S_0,
\end{equation}
with $\Delta_{\gamma}h=0$, in parallel with the harmonic structure found in the mixed time-space branch.

Moreover, using the relation $A(t)=\ln{R(t)}$, the differential equation determining the dynamical behaviour of the metric function $A(t)$ given in (\ref{mastertem}) can be written in terms of $R(t)$ as
\begin{equation}
    R\ddot{R}+\Bigl[\frac{(N-2)^2}{a^2}-1\Bigl]\dot{R}^2=0.
\end{equation}
Integrating this equation once, we get 
\begin{equation}
    R^{\frac{(N-2)^2}{a^2}-1}\dot{R}=c_1, \label{int1}
\end{equation}
where $c_1$ is an integration constant. A second integration then gives 
\begin{equation}
    R(t)=(\eta  t+c_2)^\frac{a^2}{(N-2)^2},
\end{equation}
$c_2$ is another integration constant and $\eta=\frac{(N-2)^2}{a^2}c_1$. This indicates that the metric function $R(t)$ evolves as a power law in time. This type of behaviour has been studied extensively in special cases, including FLRW cosmological backgrounds.

Finally, combining the reduced dilaton equation (\ref{lamdil}) with the temporal master equation (\ref{mastertem}), and using $A(t)=\ln{R(t)}$ and $R^{\frac{(N-2)^2}{a^2}-1}\dot{R}=c_1$, we obtain the following expression for the cosmological constant 
\begin{equation}
    \Lambda=(N-1)\Bigl [N-\frac{(N-2)^2}{a^2}\Bigl]c_1^2.
\end{equation}
We note that depending on the value of the coupling constant $a$, the cosmological constant $\Lambda$ can take negative, zero, or positive values.

To identify possible curvature singularities, we calculate the Ricci scalar
\begin{equation}
    \mathcal{R}=NH^2\Bigl[\frac{2\ddot{R}}{R}+(N-1)(\frac{\dot{R}}{R})^2 \Bigl]-R^{-2}H^{\frac{-2}{N-2}}\Bigl[\frac{2}{N-2}\frac{\Delta_{\gamma}H}{H}+\frac{N-3}{N-2}\frac{(\nabla_{\gamma}H)^2}{H^2} \Bigl],
\end{equation}
which shows that the curvature singularities could arise from the temporal part controlled by the metric function $R(t)$, or from the spatial part controlled by the metric function $H(\mathbf{x})$. A divergence in either contribution is sufficient to signal singularity in the geometry.

\subsection{Vanishing of the Cosmological Constant limit} %Vanishing Potential Limit ($\Lambda=0$)}

We now analyze the Einstein-Maxwell-dilaton theory in the absence of the Liouville potential in the action (\ref{action}). This corresponds to the limit in which the cosmological constant vanishes $\Lambda=0$. In the derivation above, the cosmological constant appears only in the remaining dilaton equation and in the temporal parts of the Einstein equations. Therefore, the algebraic constraints and spatial equations obtained earlier remain unchanged. We then impose this limit $\Lambda=0$ directly on the remaining equations.

Setting $\Lambda=0$ in the remaining dilaton equation (\ref{dlast}) gives $\ddot{A}+N\dot{A}^2=0$. Moreover, the temporal part of the $ij$ component of the Einstein equation (\ref{temij}) in this limit becomes $\ddot{A}+\frac{1}{2}\Bigl (N+\frac{(N-2)^2}{a^2}\Bigl)\dot{A}^2 =0$. Comparing these two equations, we get
\begin{equation}
    \Big( N-\frac{(N-2)^2}{a^2} \Bigl)\dot{A}^2=0 .\label{dtt}
\end{equation}
For a generic coupling constant $a^2\neq \frac{(N-2)^2}{N}$, this equation requires $\dot{A}=0$. Since $R(t)=e^{A(t)}$, the metric function $R(t)$ becomes constant $R(t)=R_0$. Substituting this result into the temporal part $\mathcal{G}_{tt}$ given in (\ref{temptt}) fixes the separation constant to vanish $\sigma=0$. However, for a nontrivial metric function $\dot{A}\neq0$, we get a special coupling constant from equation (\ref{dtt}), given as 
\begin{equation}
    a^2=\frac{(N-2)^2}{N}.
\end{equation}

For this value of the coupling constant, the remaining dilaton equation (\ref{dlast}) and the temporal part of $\mathcal{G}_{ij}$ (\ref{temij}) become identical. Therefore, the metric function $A(t)$ is not forced to be constant. Instead, the remaining temporal equation gives $ \ddot{A}+N\dot{A}^2=0$.

In terms of $R(t)=e^{A(t)}$, this equation is equivalent to  $\frac{d}{dt}(R^{N-1}\dot{R})=0$. Therefore,
\begin{equation}
    R(t)=(\eta t+c_2)^{1/N},
\end{equation}
where $\eta$ and $c_2$ are constants. Substituting this special coupling constant $ a^2=\frac{(N-2)^2}{N}$ into (\ref{temptt}) then again requires $\sigma=0$.

In the purely spatial branch $B=B(\mathbf{x})$, the vanishing cosmological constant limit always fixes the separation constant to vanish $\sigma=0$. For generic coupling, this limit also forces the metric function $A(t)$ to be constant. A nontrivial metric function $A(t)$ survives only for a special coupling $a^2=\frac{(N-2)^2}{N}$.

\section{Relation to Known Solutions and Unifying Structure} \label{sec5}

The solutions derived in this work provide a common framework for several known classes of Einstein-Maxwell and Einstein-Maxwell-dilaton geometries. The common feature of these solutions is the appearance of a conformal metric function controlled by a harmonic potential on an appropriate spatial base. This pattern appears in the static multi-center black holes, cosmological charged black hole solutions, dilatonic extensions, and nontrivial base geometries, but it is often obtained within specific choices of the base metric, coupling constants, or matter content. The derivation presented in this paper explains why this pattern arises. In the branches studied here, the Ricci flatness of the base geometry, the harmonic equation, and the conformal potential form follow from consistency of the field equations, rather than being imposed independently. This gives the solutions a broader unifying interpretation.

The first direct relation is with the static Einstein-Maxwell solutions of Majumdar and Papapetrou \cite{majumdar1947class,papapetrou1945static}. In the non-dilatonic Einstein-Maxwell limit discussed above, imposing $\Lambda=0$ forces the temporal part of the conformal potential to become constant. After absorbing the remaining scale factor into a redefinition of the time coordinate, the line element becomes
\begin{equation}
    ds^2=-[h(\mathbf{x})+S_0]^{-2}d\tau^2+[h(\mathbf{x})+S_0]^{\frac{2}{N-2}}\gamma_{ij}dx^idx^j. \label{MP}
\end{equation}

Choosing $N=3$ and the flat base geometry $\gamma_{ij}=\delta_{ij}$, the harmonic condition obtained from the field equations reduces to $\Delta_{\mathbb{R}^3}[h(\mathbf{x})+S_0]=0$. The standard four-dimensional Majumdar-Papapetrou multi-center solution is recovered by taking this harmonic function to be
\begin{equation}
    h(\mathbf{x})+S_0=U(\mathbf{x})=1+\sum_I \frac{m_I}{|\mathbf{x}-\mathbf{x}_I|},
\end{equation}
away from the source points $\mathbf{x}=\mathbf{x}_I$. The many-black-hole interpretation of these geometries was later given by Hartle and Hawking, who showed that the constants $m_I$ describe extremal charged black holes in equilibrium \cite{hartle1972solutions}. Therefore, the Majumdar-Papapetrou/ Hartle-Hawking solutions appear as the flat-base, non-dilatonic, static, zero-$\Lambda$ limit of the present construction. The generalization in the present work is that the harmonic function on flat space is replaced by a harmonic function on a Ricci flat base, and this structure is obtained from the field equations. This suggests that the present construction may provide a systematic way to search for analogous extremal charged black holes on nontrivial Ricci flat geometries, extending the usual flat base Majumdar-Papapetrou/ Hartle-Hawking setting. 

The next direct relation is with the cosmological multi-black-hole solutions of Kastor and Traschen \cite{kastor1993cosmological}. Their work extends the Majumdar-Papapetrou structure to Einstein-Maxwell theory with a positive cosmological constant and gives analytic examples of dynamical charged black holes in a de Sitter background, including the coalescence of black holes for suitable trajectories. In the non-dilatonic Einstein-Maxwell limit of the present construction with $N=3$, $\gamma_{ij}=\delta_{ij}$, and $\Lambda>0$, the field equations give the conformal potential
\begin{align}
    \Psi(\tau,\mathbf{x})=h(\mathbf{x})+\eta\tau+S_0, \qquad \Delta_{\mathbb{R}^3}h=0,
\end{align}
where the temporal contribution is tied to the nonzero cosmological constant through the temporal equations. Therefore, in this limit, the Kastor-Traschen class of solutions is recovered from the present construction. We note that in the present work, the conformal potential structure arises naturally from consistency of the field equations and was not introduced as a specialized Einstein-Maxwell ansatz.

The dilatonic black hole solutions of Gibbons-Maeda \cite{gibbons1988black} and Garfinkle-Horowitz-Strominger \cite{garfinkle1991charged} provide important early examples of Einstein-Maxwell-dilaton and string-inspired charged black hole geometries. Related cosmological dilatonic black hole solutions were also constructed in \cite{gibbons2010black, maeda2010black}. Although these solutions display related harmonic function structures, they should not all be regarded as direct limits of the present derivation, since their coupling choices, potentials, matter content, and asymptotic structures are not identical to the setting considered here. Their relevance is instead structural, as they show that the harmonic function structure of extremal charged solutions persists in dilatonic theories. By contrast, the multi-centered static solutions in dilaton-coupled Einstein-Maxwell theory constructed by Shiraishi \cite{shiraishi1993multicentered} are recovered from the vanishing-potential, static, flat-base geometry of the present solutions, after matching the conventions for the dilaton coupling and normalization of the gauge field. In this limit, the harmonic function is the standard multi-center harmonic of the Euclidean space, generalizing the Majumdar-Papapetrou class to dilatonic Einstein-Maxwell theory. in arbitrary dimensions.

The next relevant comparison is the Maki-Shiraishi construction in cosmological Einstein-Maxwell-dilaton theory \cite{10.1063/1.530167,maki1993multi}. Their model contains the same type of dilaton coupling considered here, with the dilaton coupled to both the Maxwell field and the Liouville-type cosmological term. Moreover, their solutions are interpreted as maximally charged multi-black-hole configurations on a spatially flat cosmological background. In the notation of the present work, this corresponds to taking the base geometry to be flat $\gamma_{ij}=\delta_{ij}$. The different coupling cases considered in \cite{maki1993multi} are naturally related to the Einstein-Maxwell limit, the mixed time-space branch, and the purely spatial branch derived in the present work. We emphasize that in the present work, the relevant coupling restrictions arise from the consistency of the field equations.

The present framework also clarifies the role of nontrivial Ricci-flat base geometries. We showed that the essential geometric condition is not flatness of the spatial base, but Ricci flatness. This connects the present analysis to Einstein-Maxwell and Einstein-Maxwell-dilaton solutions constructed on Eguchi-Hanson \cite{ghezelbash2017exact, ishihara2006black}, Taub-NUT \cite{fahim2025cosmological}, and Bianchi type IX geometries \cite{fahim2024exact, ghezelbash2022bianchi, fahim2021new}, which fit naturally into this structure. In these examples, the spatial base is not flat, but it is Ricci flat, and the metric function is harmonic on the base geometry, precisely as required by the field equation reduction derived above.

We summarize only a few representative examples of this relation to the literature as
\begin{equation}
\begin{aligned}
    \gamma_{ij}=\delta_{ij}, \ \phi=0, \ \Lambda=0 \Longrightarrow &\quad \text{Majumdar-Papapetrou/Hartle-Hawking}, \\
    \nonumber 
    \gamma_{ij}=\delta_{ij}, \ \phi=0, \ \Lambda>0 \Longrightarrow &\quad \text{Kastor-Traschen}, \\
    \nonumber
    \gamma_{ij}=\delta_{ij}, \ \phi\neq 0 \qquad \Longrightarrow &\quad \text{multi-center dilatonic black holes}, \\
    \nonumber
    \gamma_{ij}=\gamma_{ij}^{TN},\gamma_{ij}^{EH}, \gamma_{ij}^{BIX} \Longrightarrow &\quad \text{nontrivial base EMD/EM}.
\end{aligned}
\end{equation}

The main significance of the present derivation is structural. The comparison above explains why these apparently different constructions share the same underlying structure. We identified this structure directly from the coupled Einstein-Maxwell-dilaton equations and showed how the consistency of the field equations leads to the branch conditions, the Ricci-flatness requirement on the base geometry, and the harmonic form of the metric function.

Finally, we note that higher-dimensional interpretations of related EMD solutions have been discussed in \cite{gouteraux2011generalized, ghezelbash2017new, fahim2025cosmological}. In those constructions, special lower-dimensional EMD solutions arise from the dimensional reduction of a higher-dimensional Einstein theory coupled to a form field, where the Liouville potential is generated by the curvature of the internal space. Since the present work derives the corresponding lower-dimensional EMD structure directly from the field equations, these higher-dimensional structures may be viewed as geometric realizations of special cases within the general framework developed here.

\section{Conclusions} \label{sec6}

In this work, we studied the Einstein-Maxwell-dilaton theory in arbitrary dimensions, with the dilaton field coupled non-minimally to the Maxwell field and to a Liouville-type potential proportional to the cosmological parameter. Motivated by the repeated appearance of harmonic metric functions in exact Einstein-Maxwell and Einstein-Maxwell-dilaton solutions, we investigated whether this structure follows from the coupled field equations without prescribing the spatial base or assuming harmonic form of the metric functions.

Within the class of ansatzes considered here, we first analyzed the generic branch in which the metric function depends on both time and the spatial coordinates. We showed that the consistency of the Maxwell, dilaton, and Einstein field equations considerably constrains the allowed form of the solution. In particular, the field equations force the two dilaton couplings to be equal. Moreover, they restrict the spatial base geometry to be Ricci flat and require the metric function to be harmonic on the base geometry. After an appropriate redefinition of the time coordinate, the resulting spacetime can be written in terms of a single conformal potential. We derived an expression for the cosmological constant and showed that it can be positive, zero, or negative depending on the value of the coupling constant. Finally, we discussed special limits of the theory, including Einstein-Maxwell limit and the case in which the Liouville potential is absent.

We then considered the purely spatial branch, where the metric function depends only on the spatial coordinates. In this case, the field equations did not lead to the same equal coupling condition as in the generic branch, instead, they imposed a distinct restriction on the coupling constants. This shows that the mixed and purely spatial branches are not merely different limits of the same construction.

The main significance of this result is therefore structural. Rather than producing an isolated exact solution, the analysis identified a common field equation mechanism behind a broad class of Einstein-Maxwell and Einstein-Maxwell-dilaton geometries. In particular, the Ricci flatness of the base geometry, the harmonic nature of the metric function and the conformal potential form all arise as consequences of the coupled field equations within the present framework. This provides a unified explanation for the common structure appearing in several known solutions, including Majumdar-Papapetrou and Hartle-Hawking multi-center charged black hole geometries, the Kastor-Traschen cosmological extensions, and dynamical black holes in EMD solutions constructed on nontrivial Ricci flat bases.

Several directions remain open for the future work. It would be interesting to extend the present analysis to theories with multiple gauge fields, more general scalar potential, or less restrictive metric ansatz. Another natural direction is to construct further explicit solutions using nontrivial Ricci flat base geometries and study their global structure, horizon properties, and physical interpretation. It may also be useful to investigate rotating generalization, possible hidden symmetries, and holographic application of the resulting geometry.
\section*{Acknowledgments}
This work was supported by the Natural Sciences and Engineering Research Council of Canada.
 \section*{References}
\bibliographystyle{unsrt}
\bibliography{main}

@article{gibbons1986black,
  title={Black holes in Kaluza-Klein theory},
  author={Gibbons, Garry W and Wiltshire, David L},
  journal={Annals of Physics},
  volume={167},
  number={1},
  pages={201--223},
  year={1986},
  publisher={Elsevier}
}

@article{gibbons1988black,
  title={Black holes and membranes in higher-dimensional theories with dilaton fields},
  author={Gibbons, Gary W and Maeda, Kei-ichi},
  journal={Nuclear Physics B},
  volume={298},
  number={4},
  pages={741--775},
  year={1988},
  publisher={Elsevier}
}

@article{horowitz1991black,
  title={Black strings and P-branes},
  author={Horowitz, Gary T and Strominger, Andrew},
  journal={Nuclear Physics B},
  volume={360},
  number={1},
  pages={197--209},
  year={1991},
  publisher={Elsevier}
}

@article{aniceto2016radiating,
  title={Radiating black holes in Einstein-Maxwell-dilaton theory and cosmic censorship violation},
  author={Aniceto, Pedro and Pani, Paolo and Rocha, Jorge V},
  journal={Journal of High Energy Physics},
  volume={2016},
  number={5},
  pages={1--19},
  year={2016},
  publisher={Springer}
}

@article{hirschmann2018black,
  title={Black hole dynamics in {E}instein-{M}axwell-dilaton theory},
  author={Hirschmann, Eric W and Lehner, Luis and Liebling, Steven L and Palenzuela, Carlos},
  journal={Physical Review D},
  volume={97},
  number={6},
  pages={064032},
  year={2018},
  publisher={APS}
}

@article{yu2018cosmic,
  title={Cosmic censorship and weak gravity conjecture in the {E}instein--{M}axwell-dilaton theory},
  author={Yu, Ten-Yeh and Wen, Wen-Yu},
  journal={Physics Letters B},
  volume={781},
  pages={713--718},
  year={2018},
  publisher={Elsevier}
}

@article{goldstein2010holography,
  title={Holography of charged dilaton black holes},
  author={Goldstein, Kevin and Kachru, Shamit and Prakash, Shiroman and Trivedi, Sandip P},
  journal={Journal of High Energy Physics},
  volume={2010},
  number={8},
  pages={1--30},
  year={2010},
  publisher={Springer}
}

@article{poletti1995charged,
  title={Charged dilaton black holes with a cosmological constant},
  author={Poletti, Steve J and Twamley, Jason and Wiltshire, David L},
  journal={Physical Review D},
  volume={51},
  number={10},
  pages={5720},
  year={1995},
  publisher={APS}
}

@article{poletti1994global,
  title={Global properties of static spherically symmetric charged dilaton spacetimes with a {L}iouville potential},
  author={Poletti, Steve J and Wiltshire, David L},
  journal={Physical Review D},
  volume={50},
  number={12},
  pages={7260},
  year={1994},
  publisher={APS}
}

@article{charmousis2009einstein,
  title={{E}instein-{M}axwell-Dilaton theories with a {L}iouville potential},
  author={Charmousis, Christos and Gout{\'e}raux, Blaise and Soda, Jiro},
  journal={Physical Review D—Particles, Fields, Gravitation, and Cosmology},
  volume={80},
  number={2},
  pages={024028},
  year={2009},
  publisher={APS}
}

@article{rocha2018self,
  title={Self-similarity in {E}instein-{M}axwell-dilaton theories and critical collapse},
  author={Rocha, Jorge V and Toma{\v{s}}evi{\'c}, Marija},
  journal={Physical Review D},
  volume={98},
  number={10},
  pages={104063},
  year={2018},
  publisher={APS}
}

@article{ghezelbash2015cosmological,
  title={Cosmological solutions in five-dimensional {E}instein-{M}axwell-dilaton theory},
  author={Ghezelbash, Amir Masoud},
  journal={Physical Review D},
  volume={91},
  number={8},
  pages={084003},
  year={2015},
  publisher={APS}
}

@inproceedings{papapetrou1945static,
  title={A static solution of the equations of the gravitational field for an arbitary charge-distribution},
  author={Papapetrou, Achilleus},
  booktitle={Proceedings of the Royal Irish Academy. Section A: Mathematical and Physical Sciences},
  volume={51},
  pages={191--204},
  year={1945},
  organization={JSTOR}
}

@article{majumdar1947class,
  title={A class of exact solutions of {E}instein's field equations},
  author={Majumdar, Sudhansu Datta},
  journal={Physical Review},
  volume={72},
  number={5},
  pages={390},
  year={1947},
  publisher={APS}
}

@article{hartle1972solutions,
  title={Solutions of the {E}instein-{M}axwell equations with many black holes},
  author={Hartle, James B and Hawking, Stephen W},
  journal={Communications in Mathematical Physics},
  volume={26},
  number={2},
  pages={87--101},
  year={1972},
  publisher={Springer}
}

@article{kastor1993cosmological,
  title={Cosmological multi-black-hole solutions},
  author={Kastor, David and Traschen, Jennie},
  journal={Physical Review D},
  volume={47},
  number={12},
  pages={5370},
  year={1993},
  publisher={APS}
}

@article{maki1993multi,
  title={Multi-black hole solutions in cosmological {E}instein-{M}axwell-dilaton theory},
  author={Maki, Takuya and Shiraishi, Kiyoshi},
  journal={Classical and Quantum Gravity},
  volume={10},
  number={10},
  pages={2171--2178},
  year={1993}
}

@article{fahim2025cosmological,
  title={Cosmological dynamical black hole solutions},
  author={Fahim, Bardia H and Ghezelbash, Amir Masoud},
  journal={International Journal of Modern Physics A},
  volume={40},
  number={06},
  pages={2550008},
  year={2025},
  publisher={World Scientific}
}

@article{ghezelbash2017exact,
  title={Exact solutions to {E}instein--{M}axwell theory on {E}guchi--{H}anson space},
  author={Ghezelbash, Amir Masoud and Kumar, V},
  journal={International Journal of Modern Physics A},
  volume={32},
  number={17},
  pages={1750098},
  year={2017},
  publisher={World Scientific}
}

@article{fahim2024exact,
  title={Exact dynamical black hole solutions in five or higher dimensions},
  author={Fahim, Bardia H and Ghezelbash, Amir Masoud},
  journal={The European Physical Journal C},
  volume={84},
  number={8},
  pages={837},
  year={2024},
  publisher={Springer}
}

@article{ghezelbash2022bianchi,
  title={Bianchi {IX} geometry and the {E}instein--{M}axwell theory},
  author={Ghezelbash, Amir Masoud},
  journal={Classical and Quantum Gravity},
  volume={39},
  number={7},
  pages={075012},
  year={2022},
  publisher={IOP Publishing}
}

@article{fahim2021new,
  title={New class of exact solutions to {E}instein--{M}axwell-dilaton theory on four-dimensional {B}ianchi type {IX} geometry},
  author={Fahim, Bardia H and Ghezelbash, Amir Masoud},
  journal={The European Physical Journal C},
  volume={81},
  number={7},
  pages={587},
  year={2021},
  publisher={Springer}
}

@article{shiraishi1993multicentered,
  title={Multicentered solution for maximally charged dilaton black holes in arbitrary dimensions},
  author={Shiraishi, Kiyoshi},
  journal={Journal of mathematical physics},
  volume={34},
  number={4},
  pages={1480--1486},
  year={1993},
  publisher={American Institute of Physics}
}

@article{gibbons2010black,
  title={Black holes in an expanding universe},
  author={Gibbons, Gary W and Maeda, Kei-ichi},
  journal={Physical review letters},
  volume={104},
  number={13},
  pages={131101},
  year={2010},
  publisher={APS}
}

@article{maeda2010black,
  title={Black hole in the expanding universe with arbitrary power-law expansion},
  author={Maeda, Kei-ichi and Nozawa, Masato},
  journal={Physical Review D—Particles, Fields, Gravitation, and Cosmology},
  volume={81},
  number={12},
  pages={124038},
  year={2010},
  publisher={APS}
}

@article{garfinkle1991charged,
  title={Charged black holes in string theory},
  author={Garfinkle, David and Horowitz, Gary T and Strominger, Andrew},
  journal={Physical Review D},
  volume={43},
  number={10},
  pages={3140},
  year={1991},
  publisher={APS}
}

@article{gouteraux2011generalized,
  title={Generalized holographic quantum criticality at finite density},
  author={Gout{\'e}raux, Blaise and Kiritsis, Elias},
  journal={Journal of High Energy Physics},
  volume={2011},
  number={12},
  pages={36},
  year={2011},
  publisher={Springer}
}

@article{ghezelbash2017new,
  title={New class of exact solutions in {E}instein-{M}axwell-dilaton theory},
  author={Ghezelbash, Amir Masoud},
  journal={Physical Review D},
  volume={95},
  number={6},
  pages={064030},
  year={2017},
  publisher={APS}
}

@article{harrison1968new,
  title={New Solutions of the {E}instein-{M}axwell Equations from Old},
  author={Harrison, B Kent},
  journal={Journal of Mathematical Physics},
  volume={9},
  number={11},
  pages={1744--1752},
  year={1968},
  publisher={American Institute of Physics}
}

@article{ida2007cosmological,
  title={Cosmological black holes on {T}aub--{NUT} space in five-dimensional {E}instein--{M}axwell theory},
  author={Ida, Daisuke and Ishihara, Hideki and Kimura, Masashi and Matsuno, Ken and Morisawa, Yoshiyuki and Tomizawa, Shinya},
  journal={Classical and Quantum Gravity},
  volume={24},
  number={13},
  pages={3141--3149},
  year={2007}
}

@article{butler2019minimal,
  title={Minimal surfaces and generalized {E}instein--{M}axwell-dilaton theory},
  author={Butler, Michael and Ghezelbash, A Masoud},
  journal={International Journal of Modern Physics A},
  volume={34},
  number={12},
  pages={1950061},
  year={2019},
  publisher={World Scientific}
}

@article{ishihara2006black,
  title={Black holes on {E}guchi-{H}anson space in five-dimensional {E}instein-{M}axwell theory},
  author={Ishihara, Hideki and Kimura, Masashi and Matsuno, Ken and Tomizawa, Shinya},
  journal={arXiv preprint hep-th/0607035},
  year={2006}
}

@article{10.1063/1.530167,
    author = {Shiraishi, Kiyoshi},
    title = {Multicentered solution for maximally charged dilaton black holes in arbitrary dimensions},
    journal = {Journal of Mathematical Physics},
    volume = {34},
    number = {4},
    pages = {1480-1486},
    year = {1993},
    month = {04},
    issn = {0022-2488},
    doi = {10.1063/1.530167},
    url = {https://doi.org/10.1063/1.530167},
    eprint = {https://pubs.aip.org/aip/jmp/article-pdf/34/4/1480/19324411/1480_1_online.pdf},
}
\end{document}